\documentclass{article}
\usepackage{url}
\usepackage{listings}
\usepackage{multicol}
\usepackage{authblk}
\usepackage{multirow}
\usepackage{graphicx}
\usepackage{hyperref}
\usepackage{fullpage}
\usepackage{color}

\title{Evolving system bottlenecks in the as a service cloud}

\author{Shaun C. D'Souza}
\affil{CTO Office, Wipro Limited}
\affil{shaun.dsouza1@wipro.com}
\date{}

\begin{document}
\maketitle

%\section{Appreciation of Functionality}
\begin{abstract}

The web ecosystem is rapidly evolving with changing business and functional models. Cloud platforms are available in a SaaS, PaaS and IaaS model designed around commoditized Linux based servers. 10 billion users will be online and accessing the web and its various content. The industry has seen a convergence around IP based technology. Additionally, Linux based designs allow for a system wide profiling of application characteristics. The customer is an OEM who provides Linux based servers for telecom solutions. The end customer will develop business applications on the server. Customers are interested in a latency profiling mechanism which helps them to understand how the application behaves at run time. The latency profiler is supposed to find the code path which makes an application block on I/O, and other synchronization primitives. This will allow the customer to understand the performance bottleneck and tune the system and application parameters.
\end{abstract}

\section{Introduction}

{\bf Web N.}
Open source has enabled the development of more efficient internet systems. As application performance is a top constraint, profiling is used to verify the performance of a multi-process and multi-threaded workloads. Additionally, a good developer always makes the optimal use of the platform architecture resources. Applications are deployed today in cloud environment and as a developer it is key to ensure that the code is well documented including maintenance of architectural UML diagrams to ensure minimal errors. Cloud computing combined with service-oriented architecture (SOA) and parallel computing have influenced application software development. Code is implemented in a variety of frameworks to ensure performance including OpenMP, MPI, MapReduce \cite{dean2008mapreduce} and Hadoop. Parallel programs present a variety of challenges including race conditions, synchronization and load balancing overhead. 

{\bf Virtualization.}
This hosted infrastructure is complemented by a set of virtualization utilities that enable rapid provisioning and deployment of the web infrastructure in a distributed environment. Virtualization abstracts the underlying platform from the OS enabling a flexible infrastructure. Developers can partition resources across multiple running applications to ensure maximum utilization of resources. Additionally, systems can be ported across platforms with ease. Virtualization enables isolation of application run-time environment including kernel libraries ensuring that multi-threaded and multi-process workloads run in a scalable manner without errors. It decouples product development platforms from the conventional SDLC models enabling existing infrastructure to scale in available resources. The virtualization layer enables application profiling and development.

{\bf Open Source.}
Additionally, the cloud ecosystem is supported by the Open Source community enabling an accelerated scale of development and collaboration \cite{mahmood2013software}. This has been enabled by the internet and version control systems.

OOP and Java have enabled enterprise system architecture. Java is an algorithms, web and enterprise centric programming language. It allows for deployment of applications on a host of platforms running a virtual machine. 3 billion mobile devices run Java. Enterprise applications provide the business logic for an enterprise. Architectures have evolved from monolithic systems, to distributed tiered systems, to Internet connected cloud systems today.

Computing and the internet are more accessible and available to the larger community. Machine learning \cite{d2018system} has made extensive advances with the availability of modern computing. It is used widely in Natural Language Processing, Speech Recognition and Web Search.

\begin{itemize}
\item Web N, 10 Billion users, Intelligent machines, Turing test
\item Social media, enterprise mobility, data analytics and cloud
\item Technology and enterprise
\item Virtualization, Open Source
\item Machine learning, compilers, algorithms \cite{d2017parser}, systems
\end{itemize}

\section{Queries}

We evaluate the customer application in the internet. We developed a full-system simulator to evaluate web workloads in a server client environment including network. We found opportunity for the use of virtualization technology to efficiently utilize cloud resources. Additionally, use the virtualization layers for performance related benchmarking functionality. 

Review the use of profiling tools with the customer including COTS commercial solutions and alternative open source solutions. Both tools present a set of tradeoffs which would vary in the customer application. A commercial solution like Intel VTune, VMware vmkperf would be suitable for a range of applications and provide more accurate profiling data on a host platform running Windows.

However, as part of our solution implementation we will be reviewing the use of an open source solution in Linux perf and Xenoprof (OProfile). We would review the nature of the customer application including availability of source code. Additionally, we would assess the availability of a high level software architecture specification including UML, use case diagrams. These would allow us to evaluate a first-order survey of the system application bottlenecks on the target platform. This would include the choice of application UI (presentation layer), middle layer (business logic layer) and data access (data layer).

We would evaluate the application run time environment. If the application is implemented in C++ it would support the symbol tagged Linux kernel libraries. A java application would run with a compatible Java Runtime Environment (JRE) to ensure detection of class binary symbol information. We would evaluate the possibility of using the application source code vs. benchmarking the application binary. A binary would allow us to profile application performance information. Availability of source code would allow us to profile the application in a simulation environment including addition of custom flags and debug messages to benchmark the application. It would further allow us to customize the application to improve performance on the host platform.

As internet platforms evolve towards a cloud service model, we would evaluate the application runtime environment and opportunities for hosting the application in a private, public cloud. This would ensure application performance and scalability in a deployed setting. The application would scale in the usage model and number of users on a cloud platform.

Ensure customer is aligned with current business and technology environment. Architect and design cloud applications to support multitenancy, concurrency management, parallel processing and service-oriented architecture supporting rest services.

\begin{itemize}
\item Law of accelerating returns
\item Prices and margins, competition, converging global supply and demand, evolving business models
\item Tier vs. batch processing, open source (availability of source code), language – C++, Java, security and reliability.
\item Public cloud - SaaS, PaaS, IaaS, in-house 
\item Numbers of users, usage model
\item Structured data
\item Web N, Availability of cloud computing platforms – SaaS, PaaS, IaaS. Use of virtualization infrastructure. Rapid provisioning and performance profiling of available resources
\item Simulation. Full system simulation of end-end internet. Order of magnitude (slower)
\item Type of application. Web based application hosted on a cloud computing platform.
\end{itemize}

\section{Assumptions related to functionality / requirements}

There is a host of cloud computing infrastructure deployed on Linux based platforms. Linux is open source and supports benchmarking and profiling of various applications. Additionally, it supports the use of Virtualization like Xen and VMware. We use OOP languages including Java, C++ and python. A NoSQL database MongoDB to store the profiling results data. This data is read and output to a web browser using Meteor web server. Kibana is used to store the profiling data in Elasticsearch. A dashboard is created to analyze the data in a user viewable histogram and pie chart.

\begin{itemize}
\item {\bf Web N.} The internet is increasingly accessible to more than 10 billion users. It has been designed around Internet protocols and standards. The next generation of the web will use various Semantic web technologies.
\item {\bf Cloud computing.} Rapidly commoditized infrastructure and Linux servers
\item Linux, Python, NoSQL Database MongoDB, Meteor web server, Kibana
\item {\bf Knowledge systems.} Vast repositories of structured, unstructured data
\item {\bf Efficient programming languages.} github
\end{itemize}

With the exponential growth in technology development in the recent years we find that developers have increased access to commoditized compute technology. Platforms based on commodity Linux solution are widely deployed in the enterprise. Application developers are concerned about application performance and scalability in the cloud. Application performance bottlenecks are constantly evolving in a tiered internet. They vary around system constraints limitations in the kernel functionality. However, application scalability is bounded in fundamental constraints of application development arising from a producer consumer model. The Producer Consumer or Bounded buffer problem is an example of a multi-process synchronization challenge. It forms a central role in any Operating system design that allows concurrent process activity. 

\begin{verbatim}
semaphore empty, mutex, full

function PRODUCER
    while (true) do
        WAIT(empty);
        WAIT(mutex);
        // add item
        // increment "head"
        SIGNAL(mutex);
        SIGNAL (full);
	
function CONSUMER
    while (true) do
        WAIT(full);
        WAIT(mutex);
        // remove item
        // increment "tail"
        SIGNAL(mutex);
        SIGNAL(empty);
\end{verbatim}

\begin{figure}[!h]
    \centering
	\includegraphics[width=0.6\columnwidth]{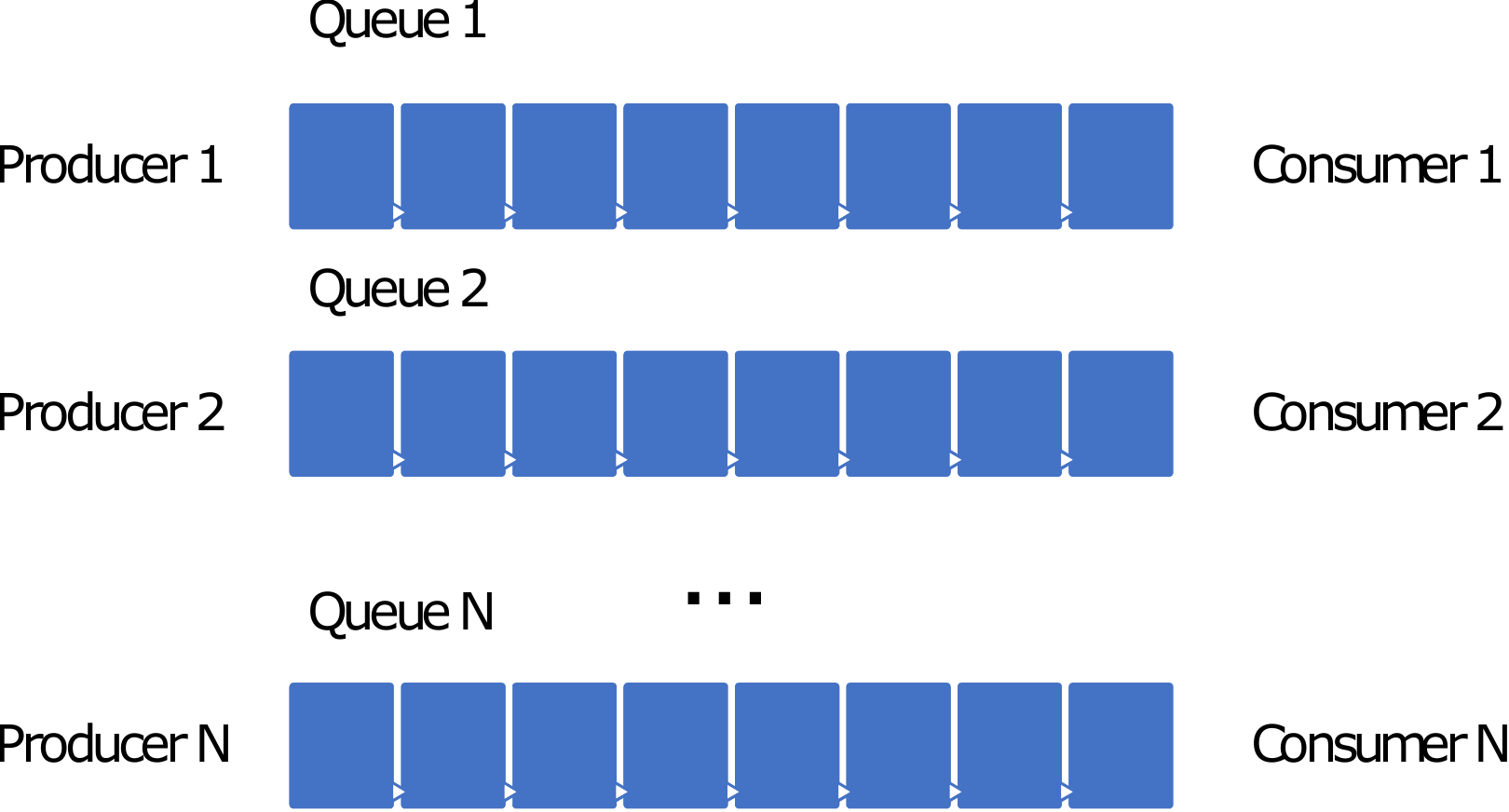}
	\caption{Producer Consumer.}
	\label{fig:prodcons}
\end{figure}

As we have N producers, N consumers and N queues in the application – Figure ~\ref{fig:prodcons} we can see that there are opportunities for the synchronization through the use of semaphores, deadlock avoidance and starvation. If we imagine infinite resources then the producer continues writing to the queue and the consumer has only to wait till there is data in the queue. The dining philosopher’s problem is another demonstration of the challenges in concurrency and synchronization.

\subsection{Application workloads today}

\begin{figure}[!h]
    \centering
	\includegraphics[width=0.2\columnwidth]{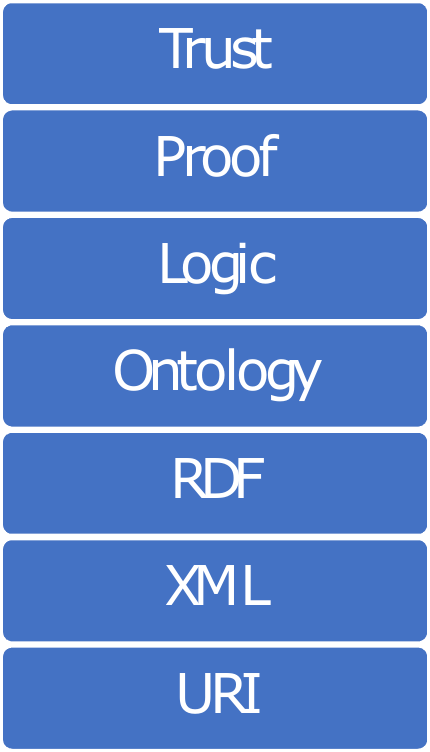}
	\caption{Semantic web stack.}
	\label{fig:semantic}
\end{figure}

In the broader context of the internet it is always beneficial to host resources close to the client consumption including providing a larger bandwidth to the consumer. Additionally, open platforms and standards enable for a balanced distribution of available bandwidth resources allowing for a scalable platform for 10 billion consumers. Innovation and advancement is enabled through open source and open platforms around internet based wireless technology. The protocol stacks comprising the future semantic web data are as – Figure ~\ref{fig:semantic}.

\section{Technical Risks \& Mitigation plan}

\begin{table}[!h]
  \centering
	\begin{tabular}{ | p{0.25\columnwidth} | p{0.65\columnwidth} | }
	\hline
	Development environment	& Availability of compilers and code generators. Availability of full-system Linux simulator supporting customer application functionality. \\ \hline
	Technology & Custom application binary support for user profiling tools including symbol tag information. Generation of dependency graphs. Recompilation of driver libraries. \\ \hline
	Availability of source code	& Annotated source code. Proprietary system applications. \\ \hline
	Customer application environment & Support for virtualization technology for Host OS and application compatibility. \\ \hline
	Product size, Business impact, Customer related	& Code reuse, Open source, Revenue, delivery deadline, New customer, customer reviews \\ \hline
	\end{tabular}
  \caption{Top 5 Risk and mitigation}
\end{table}

\section{Modeling Notation}

\begin{enumerate}
\item Unified Modeling Language (UML)
\begin{enumerate}
\item Class diagram
\item Use case
\item Sequence
\end{enumerate}
\item Eclipse 
\item JDeveloper
\item JavaScript wavi
\end{enumerate}

\begin{figure}
  \centering
  \includegraphics[width=0.8\columnwidth]{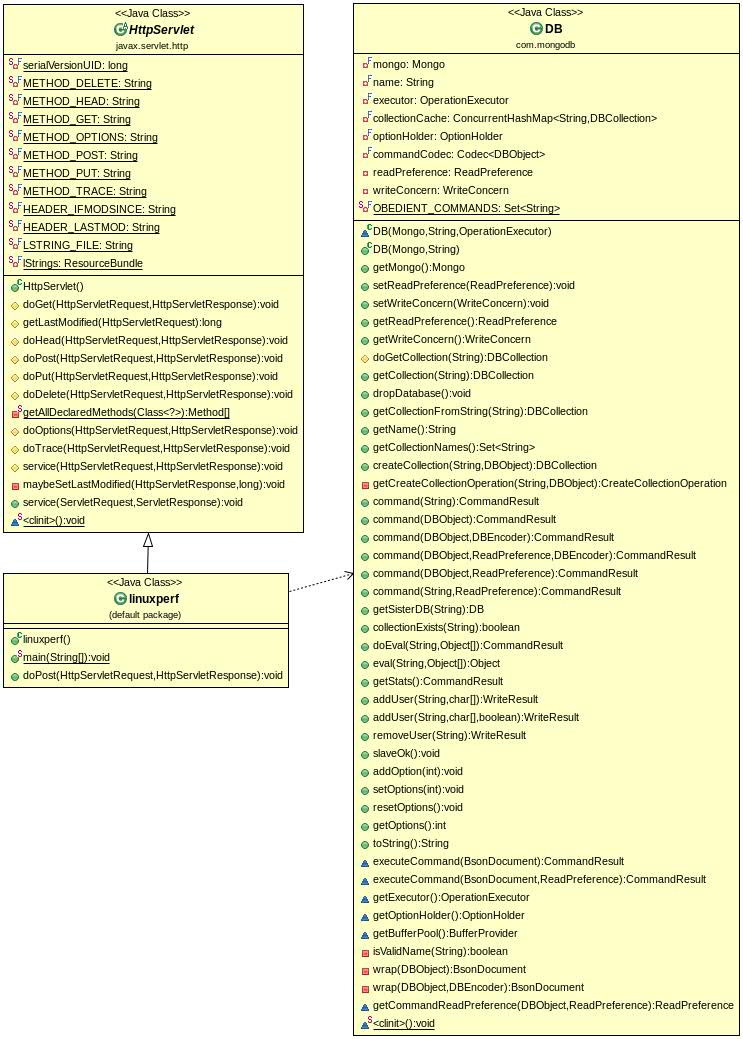}
  \caption{Tomcat MongoDB server.}
  \label{fig:tomcat}
\end{figure}

See Figure ~\ref{fig:tomcat}.

\section{Comparative Analysis}

\subsection{Alternative options considered}

There are a wide range of profiling tools for the Linux kernel.

\begin{enumerate}
\item Data collection
\begin{enumerate}
\item sysstat package – iostat, pidstat
\item sar, atop
\end{enumerate}
\item Online data - top
\begin{enumerate}
\item iotop, iftop
\end{enumerate}
\item Tracing – strace, perf\_events, mutrace, ftrace
\begin{enumerate}
\item perf-tools, gprof
\end{enumerate}
\end{enumerate}

Virtualization technology is now common in the internet datacenter. It is also used for performance gathering. Will the customer be deploying the application in a private / public cloud? Evaluate the application in the cloud infrastructure. Increasingly vendors are adopting an as a service model for hosting applications on a platform. Evaluate opportunities for improving the application performance using native and hosted virtualization techniques. As per the type of application there are a host of performance profiling environments including perf, Intel VTune, VMware vmkperf. 

System level application analysis is supported in Unix through the use of top, ps, vmstat, sar. However, these applications only allow for visibility and performance metrics at the application level. Additionally, Unix top allows the user to measure application compute and memory resource utilization. It provides information on processes running in the application.

%\subsection{Alternative options considered (contd.)}

Simulators are used to benchmark application systems. M5 \cite{binkert2011gem5} is a full-system simulator and supports modeling of compute, memory and network components. We implement an accurate view of the user system in a simulation environment. The simulator enables full system visibility of the technology stack allowing the user to configure application and kernel OS. Additionally, it is an open source simulator and can be extended to implement customer functionality to benchmark the application. It supports tracing of application and kernel function calls.

There are a host of cloud platforms available to deploy the customer application in a web environment. Cloud environment enables scalability of the application. As per the vendor environment there are a number of utilities to evaluate application performance in the cloud. 

System bottlenecks are constantly evolving. As infrastructure is increasingly being commoditized with a growth of development around open source technologies. It is essential that adequate bandwidth is provisioned in the cloud to allow for application scalability. Virtualization technology enables efficient partitioning of additional resources. As a metric, it is key to replicate scale the infrastructure maintaining redundancy to ensure quality of service in the end-to-end internet.

At the same time, it is good to ensure efficient code is developed and maintained. There are a host of projects on github that allow for efficient development of user application code.

Additionally, availability of source code enables for maintenance of code counters to measure performance. We developed a full-system simulator M5. It enables application profiling in a simulation environment including server client.

The simulator is available as open source and allows you to model compute, memory and network functionality in the system. It is an event driven simulator and supports benchmarking of key workload characteristics, check pointing, forwarding and replay. Additionally, these capabilities can be extended in the simulator ecosystem including updation of the Linux kernel to support the application code functionality and profiling. The simulator and Linux environment are accessible online. However, as we are enabling the performance profiling in simulation we do have the disadvantage that metrics gathering is at 10X slower as the entire system is running in a simulation environment.

\section{Recommended option}

As a solution we would take a multi-step approach in resolving the performance profile. As per our initial set of queries we would evaluate the availability of a high-level software architecture specification. We would evaluate opportunities for improving performance utilization and bottlenecks in the application per the specifics of the architecture. We would evaluate the application deployment in a cloud environment and support for virtualization technology. This would allow for application portability on a host of platforms and ensure scalability.

We would evaluate the application throughput requirements in the end-to-end internet. This would include all memory and network bandwidth in the system that would be utilized in a finite amount of time. We would ensure that there is no saturation of resources on all the data pipes in the platform. Data intensive workloads are usually writing large amounts of data to and from memory including serving the information to a client on the network. Consequently, as per the application requirements we would evaluate the average and peak bandwidth requirements for all the interfaces on the platform. This would ensure a first order evaluation of the application requirements.

We would then evaluate the application in Linux perf \cite{de2010new}. For the purposes of our performance profiling we will be using the Linux perf utility. Additionally, we would benchmark the application in the M5 simulator. Simulation gives us the ability to evaluate the application stack in a full system environment. We are able to view the function call trace and kernel system calls. Utilities like doxygen and gprof output a call graph for the source code statically at compile time enabling us to view function calls to locking / synchronization primitives. It outputs the control flow graph to view the structure of the program.

\section{Rationale behind suggested solution}

Perf supports performance modeling of a variety of events in the user application and kernel code. We are able to capture this data in the OS and output it to the customer in a web interface. The other set of utilities do not provide sufficient information profiling other than compute utilization and availability of memory resources. There is no capability to view thread resources in the system. Linux pthreads are scheduled and provisioned in the OS to enable performance scalability.

Utilities like gprof and doxygen support creation of UML and call graph. These enable visualization of the application function call and libraries used. They show percentage utilization of application code and opportunities for optimization.

\section{Estimation}

\begin{itemize}
\item Java KLOC per week = 0.85 
\item Python KLOC per week = 0.75
\end{itemize}

\begin{figure}[!h]
    \centering
	\includegraphics[width=0.8\columnwidth]{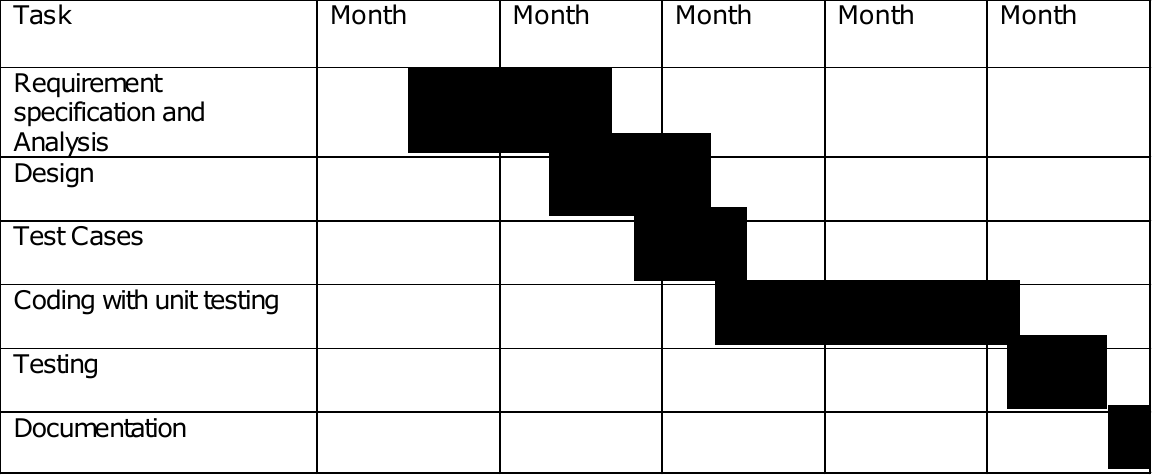}
\end{figure}

\begin{table}[!h]
  \label{tab:wbs}
  \centering
	\begin{tabular}{ | c | p{3cm} | p{1.5cm} | p{1cm} | p{2cm} | p{1cm} | p{2cm} | p{1cm} | p{1cm} | }
	\hline
	Sl. & Modules & Estimated LOC & \multicolumn{6}{|c|}{Estimated Effort (PD)} \\ \cline{4-9}
	 & & & CUT & Requirement & Design & Architecture & Testing & Total \\ \hline
	1 & Compilation and library support & 2000 & 12 & 3 & 4 & 1 & 9 & 29 \\ \hline
	2 & Code annotation & 1000 & 6 & 1 & 2 & 1 & 4 & 14 \\ \hline
	3 & Simulator integration & 1000 & 6 & 1 & 2 & 1 & 4 & 14 \\ \hline
	4 & Operating System integration & 2000 & 12 & 3 & 4 & 1 & 9 & 29 \\ \hline
	5 & Virtualization & 1500 & 9 & 2 & 3 & 1 & 7 & 22 \\ \hline
	6 & Performance analysis & 2000 & 12 & 3 & 4 & 1 & 9 & 29 \\ \hline
	7 & Input data to Elasticsearch & 1000 & 6 & 1 & 2 & 1 & 4 & 14 \\ \hline
	8 & Kibana visualization & 2000 & 12 & 3 & 4 & 1 & 9 & 29 \\ \hline
    \end{tabular}
  \caption{WBS}
\end{table}

\section{SDLC}

\subsection{SDLC model to be used}

We use an Agile, continuous development SDLC model. Incremental development is used to deliver the product in short iterations of 1 to 4 weeks. Incremental delivery includes functions and features that have been developed. Continuous integration is used to integrate work frequently. Each integration is verified and tested in an automated build to detect errors. This enables the rapid development of cohesive software.

\subsection{Rationale}

\begin{figure}[!h]
    \centering
	\includegraphics[width=0.4\columnwidth]{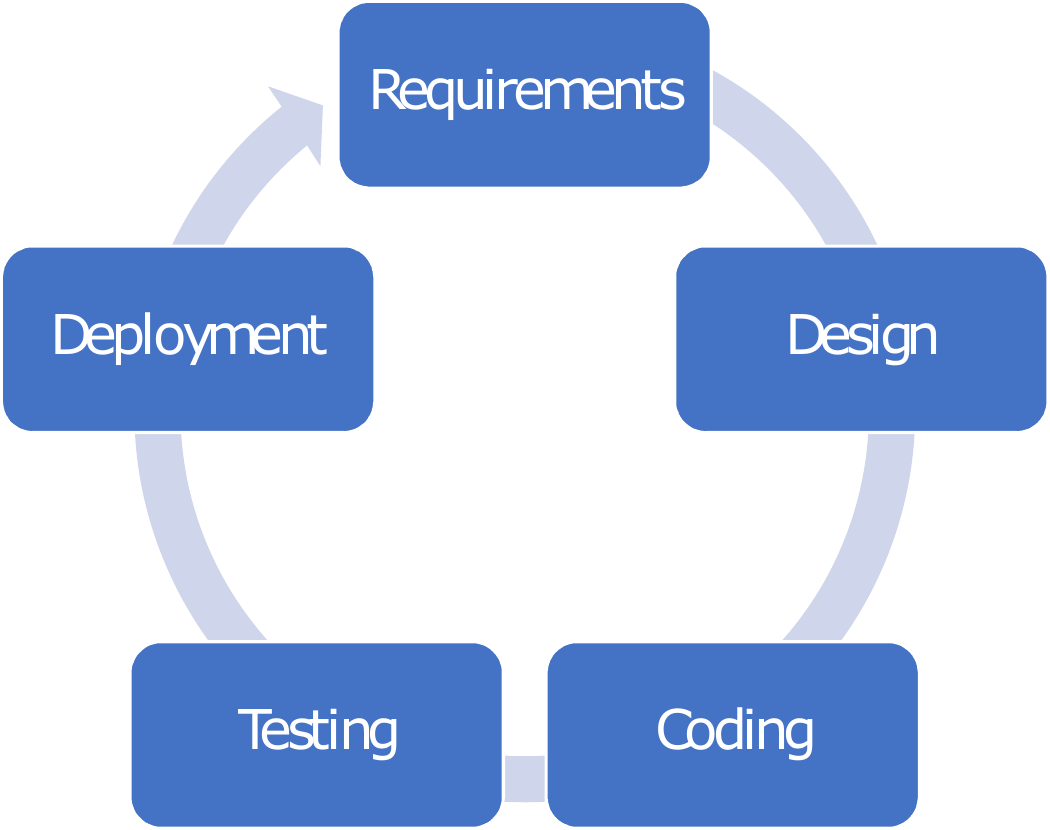}
	\caption{Agile continuous development.}
	\label{fig:sdlc}
\end{figure}

Agile development and continuous integration. Enabled early detection and integration of defect and ensures code quality. Development happens in short iterations with fully automated regression tests. High level functional requirements are documented as user stories. Software development follows a model of Figure ~\ref{fig:sdlc}.

\begin{itemize}
\item Requirements – Design – Coding – Testing – Deployment
\end{itemize}

Agile focuses on individuals and interactions, working software, customer collaboration and responding to change. Quality is maintained through the use of automated unit testing, test driven development, design patterns and code refactoring. 

\subsection{Suggested Customizations proposed}

We would use Lean techniques and Kanban. These ensure standardization and balance of the development environment.

\section{Architectural (non-functional) requirements}

\begin{itemize}
\item Interoperability
\item Scalability
\item Quality of Service
\item Time, Cost and Productivity
\item Distributed Complex Sourcing
\item Faster Delivery of Innovation
\item Increasing Complexity
\item SaaS vs. PaaS
\item Number of users, usage model
\item Reuse of Web Services
\item Agile, continuous development
\item QOS is supported in tiered cloud service provider
\end{itemize}

\section{Critical success factors}

\begin{itemize}
\item Compatibility of development environments.
\item Availability of source code and High-level architecture specification.
\item Service Oriented Architecture (SOA)
\item Linux based application hosted to run in a cloud platform.
\end{itemize}

Platforms while constantly updating themselves there is a standardization of functionality around the Linux open source movement. There has been a proliferation of development on the Linux platform and enabled commoditization of computing technology. Enterprises are now able to host a distributed platform at a marginal expense. Availability of open source software has accelerated the development of applications. Access to compiler technology in the GNU compiler collection has driven availability of shared libraries and the Linux file system. Applications can share run-time environments. Additionally, availability of version control systems has enabled collaboration across the conventional barriers. 

We consequently choose an open source solution for our performance profiler. Availability of symbol tagged application libraries enables us to view run-time statistics including kernel system calls. Applications like top and perf provide system visibility in the Linux kernel. System performance information is captured using performance counters.

\begin{itemize}
\item Quality of service
\item User base
\item Open Source, Intellectual property, Revenue, Products and services
\item Licensing, R\&D, Open innovation, open source. Strategy vary in business needs
\end{itemize}

\section{Architectural overview of the requirement and the feasibility in Linux operating system}

\begin{figure}[!h]
    \centering
	\includegraphics[width=0.3\columnwidth]{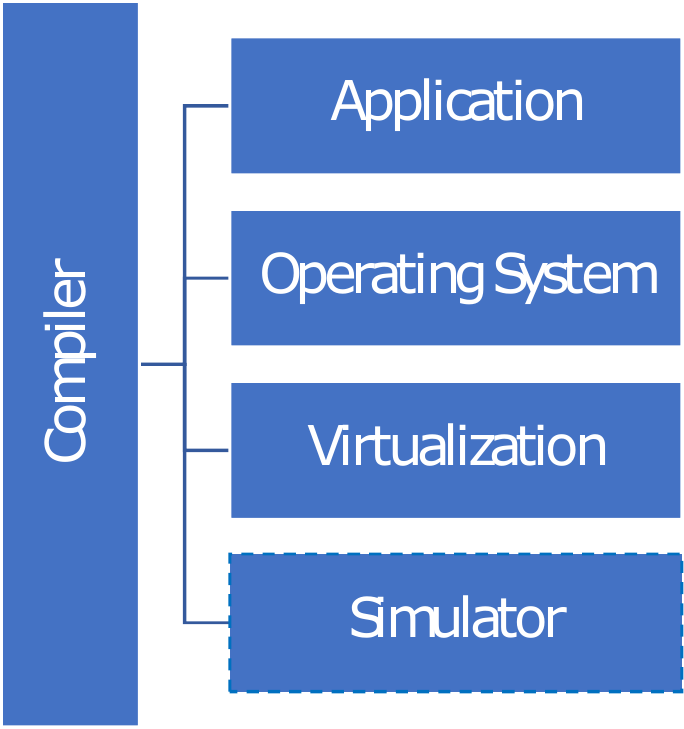}
	\caption{Host architecture.}
	\label{fig:host}
\end{figure}

Linux supports a host of virtualization and performance profiling tools \cite{du2011performance}. A virtualized architecture consists of – Figure ~\ref{fig:host}.

\begin{enumerate}
\item Data collection
\begin{enumerate}
\item sysstat package – iostat, pidstat
\item sar, atop
\end{enumerate}
\item Online data - top
\begin{enumerate}
\item iotop, iftop
\end{enumerate}
\item Tracing – strace, perf\_events, mutrace 
\item Application profiling. perf. gprof
\item Virtualization performance. VMware
\item Simulator. M5
\end{enumerate}

\begin{figure}
  \centering
  \includegraphics[width=0.6\columnwidth]{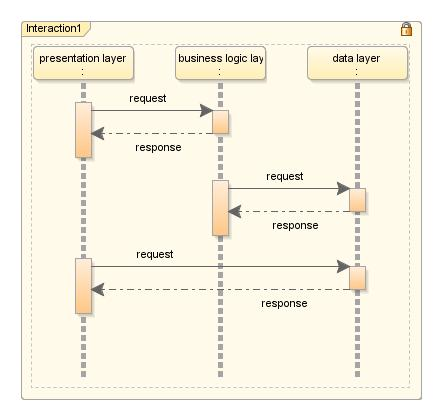}
  \caption{N-tier, SOA.}
  \label{fig:soa}
\end{figure}

The customer is interested in profiling the application on a host platform. Figure ~\ref{fig:soa} shows the characteristics of a host platform and n-tier application stack. As architects we are able to evaluate the solution using the high-level specification. An architecture use case, sequence diagram would provide us with insight on the specific usage model. We are able to investigate bottlenecks in the application. Availability of performance profiling utilities enable us to evaluate the application latency profiling requirements. 

\subsection{Linux Perf}

Perf is a performance profiling tool for Linux. It supports trace functionalities for various file system, block layer and syscall events. Tracepoints and instrumentation points are placed at logical locations in the application and kernel code. These have negligible overhead and are used by the perf command to collect information on timestamps and stack traces. It uses a set of performance counters to record and output a report for the application code.

\subsection{Intel VTune}

\begin{figure}[!h]
  \centering
  \includegraphics[width=0.8\columnwidth]{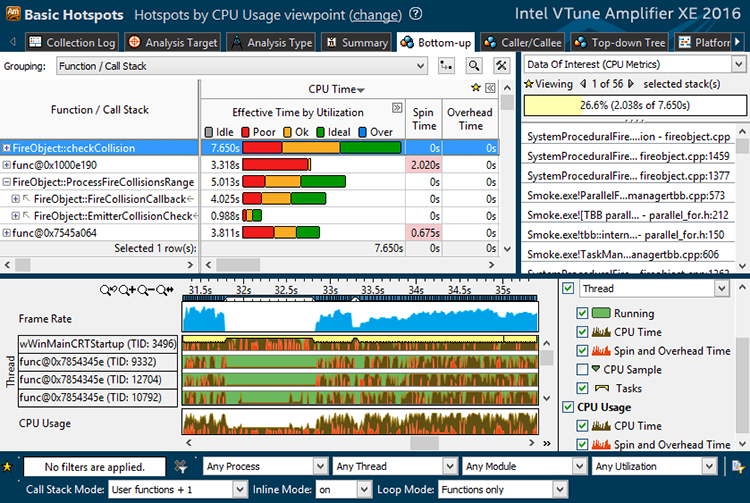}
  \caption{Intel VTune.}
  \label{fig:vtune}
\end{figure}

VTune is a proprietary performance profiling utility \cite{malladi2009using}. It supports the implementation of performance counters that are used to profile the application. It supports a GUI and command line interface. It is capable of monitoring thread level and process level performance. It supports compute performance, threading, scalability and bandwidth monitoring - Figure ~\ref{fig:vtune}.

\subsection{Simulation M5}

Simulator design and development explores a high-level architecture design space. Simulation enables the user to evaluate various deployment topologies are varying level of abstraction.  It examines the architectural building blocks in the context of performance optimization. We use the M5 architectural simulator developed at the University of Michigan, Ann Arbor. It enables us to model an end-to-end client server simulation with the native OS and application stack. There is a minimal overhead in modeling the various system components. However, the flexibility in modeling performance coupled with the high level design knowledge enable for an efficient resolution of application performance constraints.

In our research on PicoServer \cite{kgil2006picoserver}, we were able to evaluate a generic architecture for the next generation internet based on internet protocols and open standards. We proposed an architecture for Web N with 10 billion users online.

We extend this research further in proposing a generic architectural framework for performance modeling. Given an architecture specification we are able to upfront evaluate dependency graphs and bottlenecks in the design. We take various approaches from compiler development in implementing efficient scheduling algorithms for scaling application in the cloud.

\begin{figure}[!h]
  \centering
  \includegraphics[width=0.6\columnwidth]{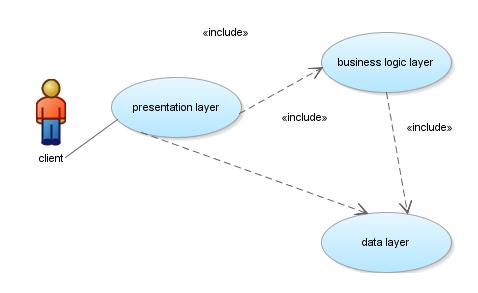}
  \caption{3-tier use case diagram.}
  \label{fig:usecase}
\end{figure}

Figure ~\ref{fig:usecase} shows the architecture of an n-tier application in the cloud. Service Oriented Architecture (SOA) applications consist of a discrete function unit that provides services to other components using a web-based protocol. A typical 3-tier application consists of a presentation layer which provides the user interface functionality. It accesses a business logic layer (middle layer) and data layer. The application is typically consolidated in a tiered cloud service provider in a SaaS, PaaS and IaaS model using a private, public cloud. 

\begin{figure}[!h]
  \centering
  \includegraphics[width=0.6\columnwidth]{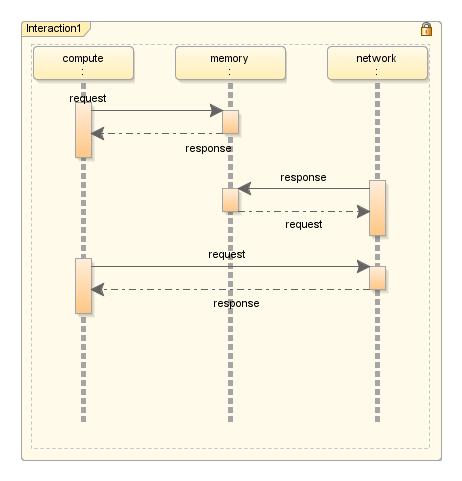}
  \caption{Web server.}
  \label{fig:webserv}
\end{figure}

In Figure ~\ref{fig:webserv}, we demonstrate the functionality of a server in the cloud which consists of compute, memory and network. With the proliferation of the internet and globalization we are seeing a commoditization of the infrastructure layer. Increasingly the server components are available as pluggable building blocks that scale in the internet. We see that a server comprises of compute and network blocks that are used to access memory. Additionally, some architectures enable direct access to network elements to improve performance and reduce wait and synchronization times.

We evaluate the opportunities for building a set of heuristics for performance profiling including ad hoc and statistical techniques and algorithms such as shortest path, directed / un-directed graphs and minimum spanning tree. Evaluate the dynamic call tree.

\section{Types of process wait state used to achieve the profiling}

Unix POSIX supports a variety of multi-threading and multi-process synchronization primitives. Pthread library is used to implement multi-threading. It supports the creation of threads using pthread\_create and synchronization wait using pthread\_join. Pthread supports the use of mutually exclusive locks through the use of pthread\_mutex\_t.

Unix supports multiprocessing capabilities through the use of fork, join. Multi-processor synchronization is implemented using wait and semaphore locks semget, semop. Additionally, there are a host of programming practices that help in the implementation of efficient parallel code with reduced wait times, starvation and deadlocks / livelocks. Linux top enables viewing of various system statistics including user / kernel mode, idle and I/O wait. Perf supports the profiling of sleep and wait time in a program. These are obtained using the perf profiling events sched.

\begin{verbatim}
>> perf record -e sched:* -g -o perf.data.raw <<application>>
>> perf inject -v -s -i perf.data.raw -o perf.data
 
>> perf report --stdio --show-total-period -i perf.data
\end{verbatim}

Latency is derived from multiple sources including scheduler, I/O and context switch. A waiting process can be blocked on events including availability of network, access to memory \cite{ghemawat2003google}. Multi-process systems enable optimization of the wait time by scheduling priority processes while a concurrent process is waiting on an external resource. 
There are various sources of latency in the Linux kernel
\begin{itemize}
\item Call to disk
\item Memory management
\item Fork and exit of a process
\end{itemize}

For an application blocked on I/O, we will see a latency on a read / write response to memory mapped input output (MMIO). Application driver code accesses a reserved section of the system memory map to communicate with external devices. Delayed access to a memory location results in an increased execution time of the driver library. MMIO’s are implemented using kernel ioread and iowrite. Address information is configured using ioremap.

perf supports profile gathering for various types of events sched, cpu-clock, syscalls, ext4, block.

\begin{verbatim}
>> perf record -e sched:* -e cpu-clock -e syscalls:* -e ext4:* -e block:* <<application>>
\end{verbatim}

\begin{figure}[!h]
    \centering
	\includegraphics[width=0.8\columnwidth]{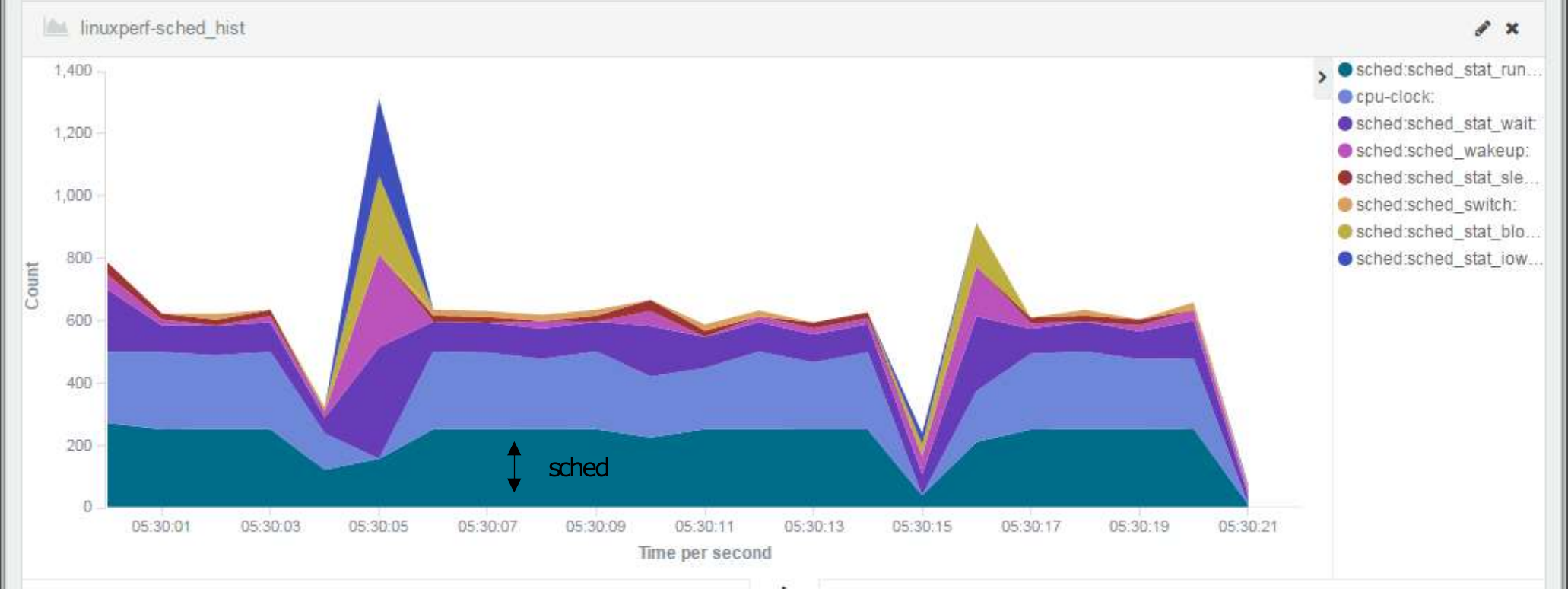}
	\caption{Multi-threaded / process use case.}
	\label{fig:multithread}
\end{figure}

We investigate the opportunities for reducing wait times in a multi-threaded, multi-process workload. We see that a large amount of time is used in sched events including pthread mutex and join - Figure ~\ref{fig:multithread}.

\subsection{Xenoprof architecture}

\begin{figure}[!h]
    \centering
	\includegraphics[width=0.6\columnwidth]{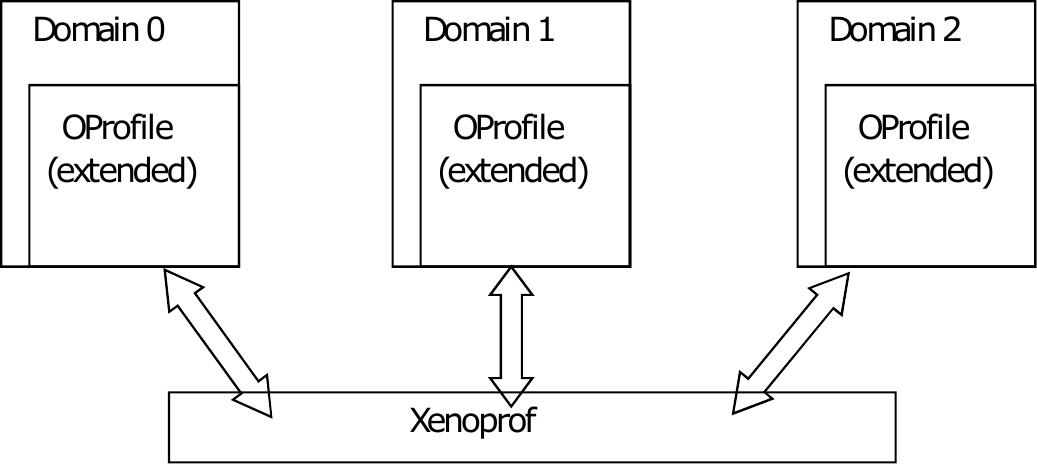}
	\caption{Xenoprof architecture.}
	\label{fig:xenoprofarch}
\end{figure}

Xenoprof is an open source profiling utility base on OProfile ~\ref{fig:xenoprofarch}. It is developed on the Xen virtual machine environment and enables the user to gather system wide data. Xen is an open source virtual machine monitor (VMM). OProfile can be used to profile kernel and user level applications and libraries. It enables profiling of applications running in a virtualized environment. 

\begin{lstlisting}[caption=xenoprof output, label=xenoprof]
Function	%Instructions	Module
e1000_intr	13 .32	e1000 
tcp_v4_rcv	8 .23	vmlinux 
main	5 .47	rcv22
\end{lstlisting}

Virtualization profiling mechanisms provide real time capabilities vs. a simulator. As all the application system calls are serviced in the virtualization engine it is suitable to extend the VMM to support profiling. This functionality enables gathering of system wide data. 

\begin{figure}[!h]
    \centering
	\includegraphics[width=0.4\columnwidth]{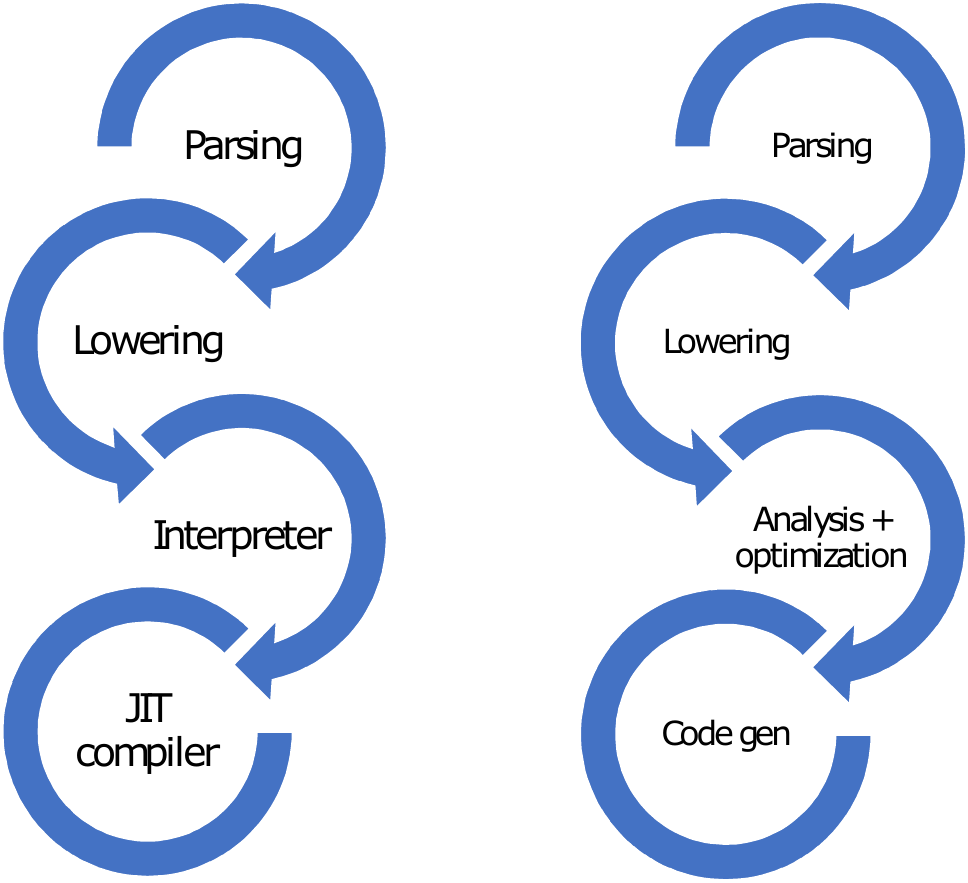}
	\caption{V8 JS, g++ compiler.}
	\label{fig:v8js}
\end{figure}

High level architecture specification enables us to define a dependency graph for the application. This is facilitated by the availability of source code. A compiler is a sequential batch architecture - Figure ~\ref{fig:v8js}. AST represents the structure of the source code. Parser turns flat text into a tree. AST is a kind of a program intermediate form (IR).  

{\bf Compilers and translators.} Compilers translate information from one representation to another. Most commonly, the information is a program. Compilers translate from high-level source code to low-level code. Translators transform representations at the same level of abstraction.

\begin{itemize}
\item Windows – 50 million LOC
\item Google internet services – 2 billion LOC
\end{itemize}

Compilers enables us to evaluate the application dependency graphs at compile time. It allows us to find cyclical dependencies in the program that might cause a deadlock. At runtime we would find extremely large idle wait time consumed by a blocking thread. In such scenarios the OS scheduler would stop the process and release its resources and restart.

A number of IDE’s support the generation of a dependency graph for software profiling. MS Visual Studio generates a dependency for the Project Solution. A Dynamic call tree can be used to evaluate application critical paths at architecture design time.

Additionally, programs like strace output library information and can be used to evaluate program execution paths. strace provides system wide profiling including accesses to kernel system calls. Trace utilities like strace, ltrace and dtrace can be used to profile system performance.

\begin{verbatim}
>> strace -rT
<<strace output>>
\end{verbatim}

\section{Mechanism used to get the profiling data}

\subsection{Linux perf}

Performance data metrics will be gathered using the perf utility. There are a large number of Unix programs that are available for performance profiling including top, iostat and sar. We run the perf utility in user mode to collect data for a specific application. The call-graph option is used to obtain function sub-tree information

\begin{verbatim}
>> perf record -g <<application>>
\end{verbatim}

We can profile system wide metrics using perf sleep. This enables the perf utility to run in the background while the application is running on the host platform. This approach can be used in validating system state for network and web server applications like tomcat and apache web server. perf captures performance data for various shared libraries including system kernel drivers. Users can profile drivers loaded using insmod, lsmod.

\begin{verbatim}
/lib/modules/kernel/drivers/ .ko
/lib/ .so

>> perf record -a sleep <<N>> &
<<application>>
\end{verbatim}

Recorded data is saved in a file perf.data for offline use. There a variety of utilities that enable viewing the data. We use the report implementation to view a percentage utilization of the shared libraries and functions in the application - Listing ~\ref{perf}.

\begin{verbatim}
>> perf report
\end{verbatim}

\begin{lstlisting}[caption=perf output, label=perf]
    29.88%    a.out  libstdc++.so.6.0.19  [.] std::basic_ostream<char, std::char_traits<char> >& std::__ostream_insert<char, std::char_traits<char>>(std::basic_ostream<char, std::char_traits<char> >&, char const*, long)
    17.53%    a.out  libstdc++.so.6.0.19  [.] std::basic_filebuf<char, std::char_traits<char> >::xsputn(char const*, long)            
    10.09%    a.out  libstdc++.so.6.0.19  [.] std::basic_streambuf<char, std::char_traits<char> >::xsputn(char const*, long)          
     7.60%    a.out  libstdc++.so.6.0.19  [.] std::ostream::sentry::sentry(std::ostream&)                                             
     5.86%    a.out  libc-2.19.so         [.] __memcpy_sse2_unaligned                                                                 
     5.50%    a.out  libstdc++.so.6.0.19  [.] std::basic_ostream<char, std::char_traits<char> >& std::operator<< <std::char_traits<char>>(std::basic_ostream<char, std::char_traits<char> >&, char const*)
     4.67%    a.out  [kernel.kallsyms]    [k] 0xffffffff811ee670                                                                      
     4.31%    a.out  libc-2.19.so         [.] strlen                                                                                  
     3.05%    a.out  libstdc++.so.6.0.19  [.] std::codecvt<char, char, __mbstate_t>::do_always_noconv() const                         
     2.26%    a.out  a.out                [.] _ZStlsISt11char_traitsIcEERSt13basic_ostreamIcT_ES5_PKc@plt                             
     1.66%    a.out  libstdc++.so.6.0.19  [.] strlen@plt                                                                              
     1.66%    a.out  libstdc++.so.6.0.19  [.] _ZSt16__ostream_insertIcSt11char_traitsIcEERSt13basic_ostreamIT_T0_ES6_PKS3_l@plt       
     1.54%    a.out  libstdc++.so.6.0.19  [.] memcpy@plt                                                                              
     1.07%    a.out  a.out                [.] foo()                                                                                   
     0.99%    a.out  libstdc++.so.6.0.19  [.] _ZNSo6sentryC1ERSo@plt                                                                  
     0.91%    a.out  libstdc++.so.6.0.19  [.] _ZNSt15basic_streambufIcSt11char_traitsIcEE6xsputnEPKcl@plt                             
     0.83%    a.out  a.out                [.] main                                                                                    
     0.55%    a.out  a.out                [.] bar()                                                                                   
     0.04%    a.out  libstdc++.so.6.0.19  [.] std::__basic_file<char>::xsputn_2(char const*, long, char const*, long)                 
\end{lstlisting}

\subsection{Network drivers}

We evaluate the performance of the Linux network drivers using perf events sock, net and skb. scsi information is obtained using the scsi event.
\begin{verbatim}
>> perf record -e sock:* -e net:* -e skb:* -e scsi:* -e cpu-clock <<application>>
\end{verbatim}

\subsection{gprof}

gprof is used to profile data using compiler annotated binary format. Application is compiled using the –pg option.

\begin{verbatim}
>> g++ -pg <<application.cpp>>
\end{verbatim}

After the program has been compiled it will output a gmon.out file on run containing the application performance information. This can be viewed in the gprof application.

\begin{verbatim}
>> gprof <<application>>
\end{verbatim}

\begin{lstlisting}[caption=gprof output]
  %   cumulative   self              self     total
 time   seconds   seconds    calls  ms/call  ms/call  name
 41.64      0.12     0.12                             main
 31.23      0.21     0.09        1    90.56    90.56  foo()
 26.02      0.29     0.08        1    75.47   166.02  bar()
  0.00      0.29     0.00        3     0.00     0.00  std::operator|(std::_Ios_Openmode, std::_Ios_Openmode)
  0.00      0.29     0.00        1     0.00     0.00  _GLOBAL__sub_I__Z3foov
  0.00      0.29     0.00        1     0.00     0.00  __static_initialization_and_destruction_0(int, int)
\end{lstlisting}

Simulators enables full-system visibility including application and OS. Simulation provides the user with a range of metrics related to application performance. Additionally, with the availability of simulator and application OS source code, the user can customize debug messages to be outputted in the simulator and kernel dmesg logs. This enables us to evaluate the metrics in an offline mode and profile application characteristics including function bottlenecks.

Unix mutrace is used to obtain thread mutex debug information including contention and wait time - Listing ~\ref{mutrace}. It provides information on the number of mutex locks used in the design, the number of times the lock changed and the average wait time for each lock.

\begin{lstlisting}[caption=mutrace output, label=mutrace]
Mutex #   Locked  Changed    Cont. tot.Time[ms] avg.Time[ms] max.Time[ms]  Flags
       0        8        4        4    45381.448     5672.681     6303.132 M-.?-.
\end{lstlisting}

There are a host of utilities that enable measuring network packet latency. Systemtap provides the user with the ability to define event handlers for the Linux kernel system calls. The user is able to log system calls and event timing information.

Availability of driver source code and build environment enables the user to add printk KERN\_DEBUG messages in the kernel logs. These provide timing information for the driver execution. However, the driver libraries have to be rebuilt.

\begin{figure}[!h]
    \centering
	\includegraphics[width=0.5\columnwidth]{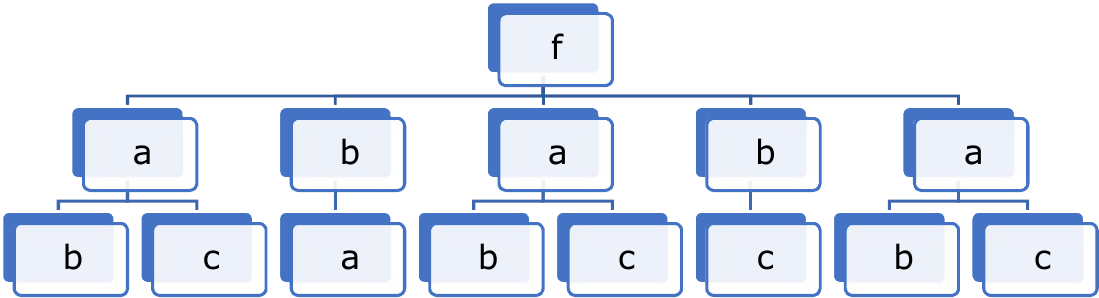}
	\label{fig:ast}
\end{figure}

\begin{figure}[!h]
    \centering
	\includegraphics[width=0.2\columnwidth]{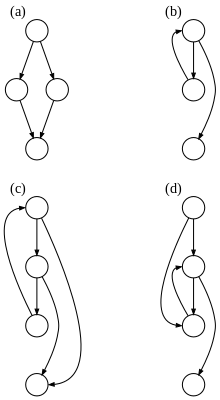}
	\caption{Control flow graph.}
	\label{fig:cfg}
\end{figure}

Compilers support the generation of an Abstract Syntax Tree (AST). This can be used to traverse multi-threaded, multi-process libraries to obtain critical path information. Availability of output from gprof and doxygen enables us to view the frequently used libraries and function calls. Compilers enable us to evaluate a static compile time profile for the application code. A control flow graph (CFG) is a graph representation of the user code. They are built recursively at compile time and are a high-level representation of the source code. 

\begin{itemize}
\item CFG = (N, E)
\item N – Set of nodes
\item E $\subseteq$ N X N – Set of edges
\end{itemize}

Nodes in the graph constitute the basic blocks in application code including the function call and libraries accessed. These are used to evaluate critical paths in the design and dependency chains. Application profiling at compile time enables the user to profile the application during architecture design. The CFG is essential to compiler optimizations and static analysis. It is used in various directed approaches to latency reduction including scheduling and parallelism.

Compilers support a variety of optimizations including dead code elimination, common sub-expression elimination, loop invariant code motion, loop unrolling and parallelization. Detection of independent sub-graphs in the application code is used in application parallelizing. Function call and for loops can be partitioned by the compiler at compile time to optimize the application code for a multi-process / distributed cloud platform. Addition of user directives in the application code enables auto-partitioning of the code to ensure performance scalability.

\section{Data storage, retrieval and presentation}

We explore the usage of various visualization features in presenting the perf report data to the customer. We use the script command to output a raw report for the performance data. This includes information on the application command, time stamp and shared library related to the event. This data is exported into an Excel csv file.

Perf supports a set of command line utilities that can be used to visualize the report data information.

\begin{verbatim}
>> perf report --call-graph --stdio
\end{verbatim}

Script is used to display and store a trace output for the test run.
\begin{verbatim}
>> perf script
\end{verbatim}

\subsection{Logstash}

\begin{figure}[!h]
  \centering
  \includegraphics[width=0.6\columnwidth]{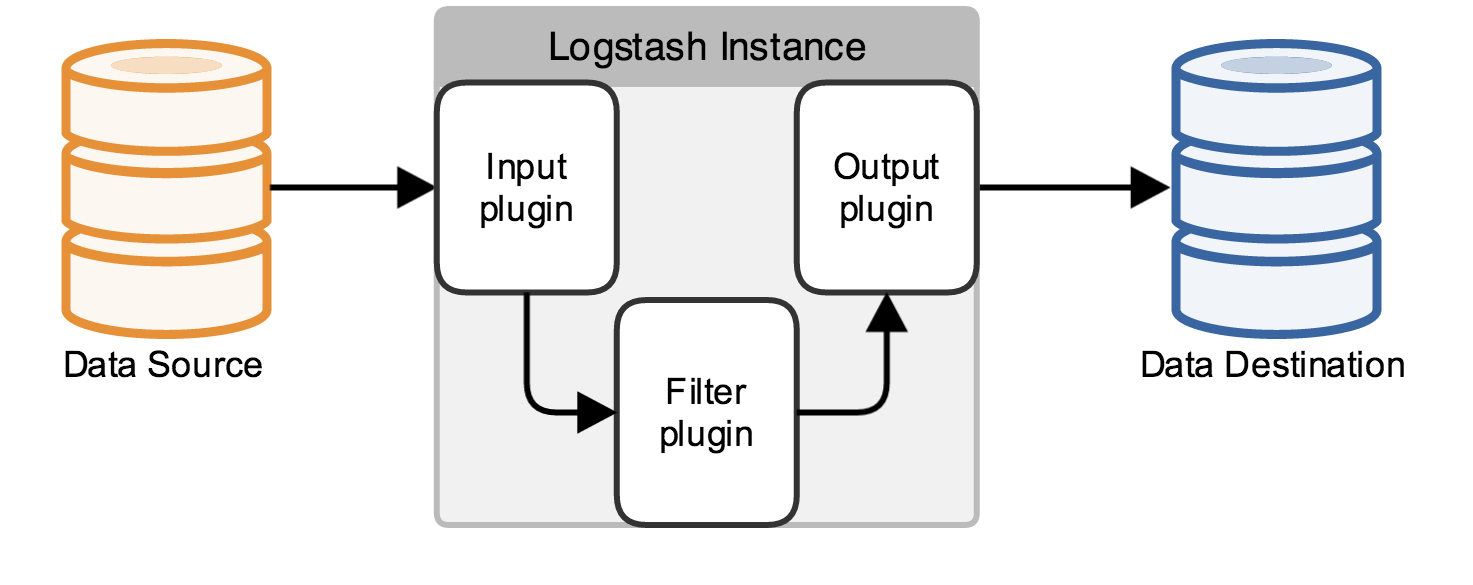}
  \caption{Logstash pipeline.}
  \label{fig:logstash}
\end{figure}

Logstash is used to input the user data into Kibana. A conf file is used to setup the Logstash pipeline – Figure ~\ref{fig:logstash}.

We use the csv input filter to read the csv data file. An output filter is used to add the data to the Elasticsearch index. A template format is specified for reading the csv column values in the date time and string format. Time values of ss.SSS are used to read the perf event logs.

\begin{verbatim}
>> cat linuxperf.csv | logstash -f linuxperf.conf
\end{verbatim}

\subsection{Elasticsearch}

Elasticsearch is a distributed and scalable data store. It provides search and analytics capabilities with Rest services. It supports indexing of a JSON document. Input is through Logstash.

\subsection{Kibana}

The data set is loaded in Kibana and a visualization dashboard is created to display the performance data to the customer.

\begin{itemize}
\item Data output in json format
\item Stored in Kibana Elasticsearch
\item Input using Logstash
\end{itemize}

\begin{verbatim}
>> curl 'localhost:9200/_cat/indices?v'
>> curl -XDELETE 'http://localhost:9200/linuxperf'
>> curl -XPUT localhost:9200/_bulk --data-binary @linuxperf.json
\end{verbatim}

Availability of user data in a csv format enables us to store the data in a NoSQL database like MongoDB. Documents are stored in the database in the json format consisting of field value pairs. MongoDB provides high performance data persistence. A collection is created for the perf data. Data is then input to the database using the pymongo library. Users can insert, update documents in the database using json dictionary values.

Alternatively, the Linux perf data is stored in the run folder in the perf.data file. This is used alongside a host of perf report utilities to view the application call graph and utilization information.

gprof similary supports output in the gmon.out format. Applications are compiled in g++ for profiling using the symbol table information in the application binary. Run-time application call graph information is stored in the gmon.out file for profiling. gprof profiles function execution utilization information.

\section{Characteristics of the user interface and the details about its functionality}

\begin{figure}[!h]
  \centering
  \includegraphics[width=0.8\columnwidth]{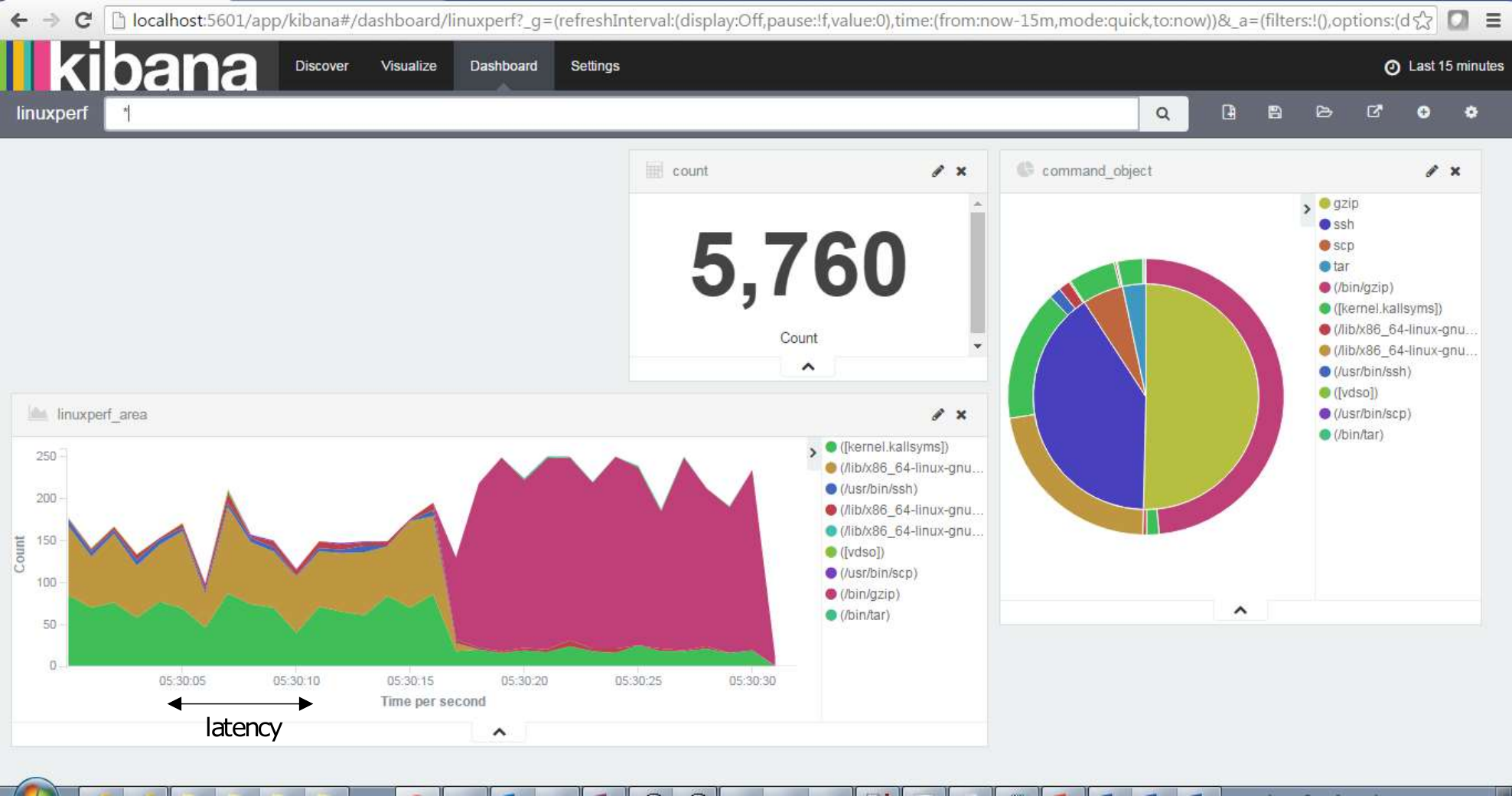}
  \caption{Kibana dashboard.}
  \label{fig:dashboard}
\end{figure}

Kibana is an open source framework for data visualization \cite{gupta2015kibana}. It enables the user to analyze and explore structured and unstructured data in an intuitive interface. It supports the use of graphs, histogram and pie charts on large amounts of data. Users can view and discover the data in the UI. They can create various visualizations that can be integrated into a comprehensive dashboard. Figure ~\ref{fig:dashboard} shows a custom dashboard implemented for the Linux perf data.

\subsection{Kibana user interface visualization}

We evaluate a network workload consisting of a secure file transfer and zip compression of the data. As can be seen in the data a large amount of time is utilized in the libcrypto and gzip application. The histogram plots the total number of events per second as recorded / reported in perf. We additionally output a pie chart of the percentage of utilization for each application and the shared libraries accessed at runtime. We capture the data for all resources on the system for the duration of the application.

As the data is input in Kibana using Logstash, we create an index for the loaded data. The data can now be viewed as a list of input values for discovery. We use this data to create a set of visualizations. We create an area plot for the performance data using the logged time values in ss.SSS. We use the date histogram feature on the X-axis and split the plot using the run time command information. We create a pie chart using the command and corresponding shared library information in a sub-plot. Time series values are encoded in the date number format. Other input data is input in the string format with the option `not\_analyzed' set.

\begin{figure}[!h]
  \centering
  \includegraphics[width=0.8\columnwidth]{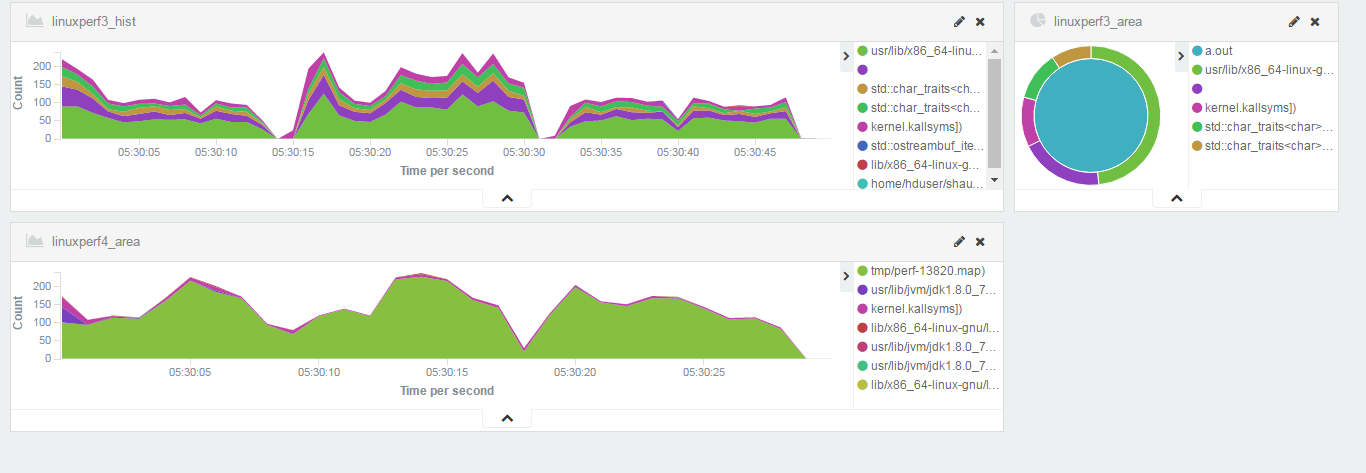}
  \caption{Kibana visualization.}
  \label{fig:visual}
\end{figure}

\subsection{Python, MongoDB, Tomcat server}

\begin{figure}[!h]
  \centering
  \includegraphics[width=0.8\columnwidth]{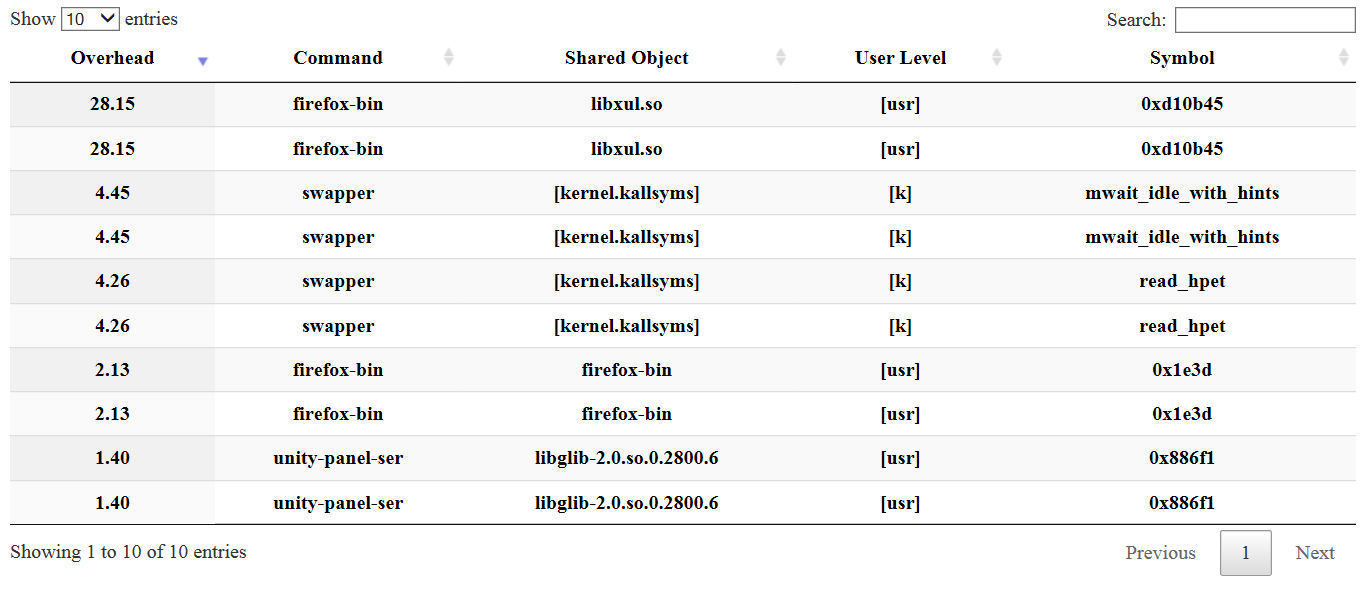}
  \caption{HTML DataTable.}
  \label{fig:html}
\end{figure}

We additionally investigate the use of a NoSQL database in storing the results data. This is an architecture deployed in the Holmes Helpdesk platform. Python is used to update the performance profiling data in MongoDB code. This is then output to the customer in a HTML file in a DataTable - Figure ~\ref{fig:html}.

{\bf HTML DataTable.}
JavaScript DataTables code runs natively on the HTML file allowing the user to browse, sort the performance data.
A Tomcat server is enabled to output the MongoDB performance data to a Java Servlet so the customer can access the information remotely - Figure ~\ref{fig:tomcat}. 

\section{Functional and nonfunctional requirements of the profiler}

The performance profiler has been developed for a Linux platform and supports the Unix POSIX function calls. It should support the C++, Java development libraries and run-time environments. There should be a compatible JRE running on the system. Additionally, a lot of the UI features will be supported in a browser based application. The platform features like Kibana and DataTables will use a browser.

\begin{itemize}
\item The application must support the linux-common-tools corresponding to the kernel release.
\item Availability of High-level architecture documents. 
\item Requirements engineering – Implementation – Testing – Evaluation
\end{itemize}

We will be exploring the design space of a Linux performance profiler that would allow us to evaluate bottlenecks arising from an inefficiently designed application. The application would support an intuitive user interface and present a metrics dashboard to the user highlighting the compute utilization for the application code. The user would be able to view the total run time for the application and library wise break down of run-time utilization. We would design the user interface such that any user would be able to evaluate the performance characteristics of an application. 

The profiler would run in a Linux based environment enabling the reuse of existing Unix kernel and application libraries. There are a number of Unix utilities that support application and system level performance monitoring. For the purposes of our investigation we will explore a hybrid of open source and COTS implementation including a simulator and compiler design. Design of the simulator and compiler are highly intensive projects with simulator designs running into 100K+ LOC and compilers at 50K+ LOC. As we will be exploring open source alternatives in our design. We will look at the use of the M5 simulator which is a full-system architectural simulator and the Unix g++ and java compilers.

Unix standards enable us to run a host of application on a shared platform architecture. Additionally, with the availability of open libraries and kernel binaries the user is able to observe the performance of an application in the Linux environment. Utilities like top and iostat enable us to view system compute utilization and I/O wait times. We will assume the functionality of an SOA application running on a web-based platform. The application would run on the OS natively enabling us to benchmark the application and gather data for offline viewing. We would provide all the application utilities on the test platform. The user would launch the application using a script which would gather the data and input it to Kibana using Logstash. We would run Kibana and Elasticsearch on a user machine for viewing the results data. It is possible that running the data viewer utility on the same test platform could corrupt the data. Alternatively, we could run the Kibana UI on a VMM on the same physical machine. 

\begin{itemize}
\item Non-functional requirements include data requirements, constraints and quality requirements.
\item Product requirements – portability, reliability, usability, efficiency, performance
\item Organization – delivery, implementation, standards
\item Availability of documented code. 
\end{itemize}

Eclipse is an open source IDE and supports a variety of programming languages including plugin functionality. Eclipse supports the standard GNU environment for compiling, building and debugging applications. The CDT is a plugin which enables development of C/C++ applications in Eclipse. It enables functionality including code browsing, syntax highlighting and code completion.

Eclipse supports a number of programming languages including C/C++, Java, PHP, XML, and HTML. It is an open source IDE and can be used on multiple platforms including Windows, Linux. It supports plugins to extend the functionality of the IDE for source code language modeling and analysis.

\begin{figure}[!h]
  \centering
  \includegraphics[width=0.6\columnwidth]{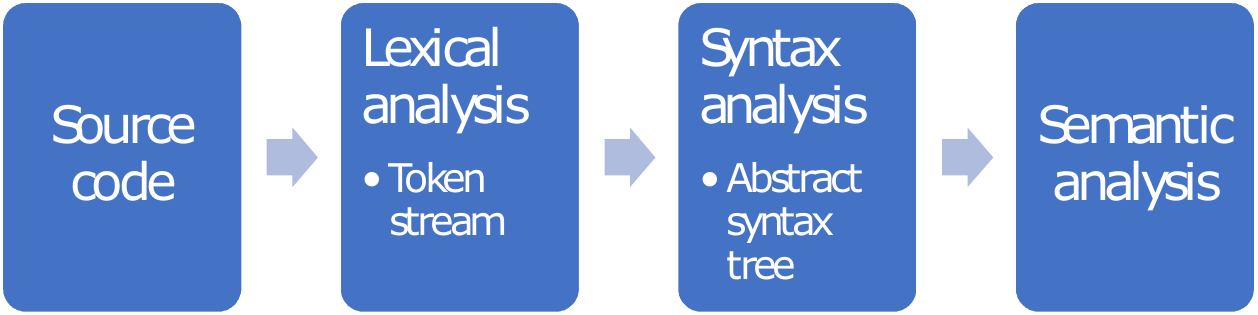}
  \caption{Stages of compilation parser.}
  \label{fig:compiler}
\end{figure}

We use the CDT to function as a compiler frontend - Figure ~\ref{fig:compiler}. The CDT uses a translation unit to represent a source file cpp and h. The CDT core supports a Visitor API which is used to traverse the AST. AST rewrite API is used to update the source code. We access the code AST using the Eclipse CDT API.

We evaluate the generation of a dependency graph for software profiling. The AST is used to profile critical application paths including access to lock and synchronization primitives. These enable us to evaluate a latency profile at design time. Eclipse supports a plugin to analyze library dependencies. MS Visual Studio generates a dependency graph for all the drivers and files in the solution architecture. 

\section{Conclusion}

We have looked at architectures for the next generation enterprise including end to end solutions for the web infrastructure. These highlight the challenges in bringing billions of users online on a commodity platform. There is a large opportunity in enabling technology consumption for more than a billion users. Technologies like social media, enterprise mobility, data analytics and cloud are disrupting the enterprise. Research and Development (R\&D) is an enabler for the enterprise. We are in the midst of a modern day information revolution enabled by increased availability of technology. I have highlighted the technology direction and challenges that we as architects foresee and resolve. These include researching machine learning, compilers, algorithms and developing efficient language models to enable applications and systems in the enterprise.

\bibliographystyle{abbrv}
\bibliography{bibliography-file} 

\end{document}